**Terahertz quantum cascade lasers with thin resonant-phonon depopulation active regions and surface-plasmon waveguides**

M. Salih,[a] P. Dean, A. Valavanis, S. P. Khanna, L. H. Li,

J. E. Cunningham, A. G. Davies, and E. H. Linfield

School of Electronic and Electrical Engineering, University of Leeds,

Leeds LS2 9JT, UK

We report three-well, resonant-phonon depopulation terahertz quantum cascade lasers with semi-insulating surface-plasmon waveguides and reduced active region (AR) thicknesses. Devices with thicknesses of 10, 7.5, 6, and 5 µm are compared in terms of threshold current density, maximum operating temperature, output power and AR temperature. Thinner ARs are technologically less demanding for epitaxial growth and result in reduced electrical heating of devices. However, it is found that 7.5-µm-thick devices give the lowest electrical power densities at threshold, as they represent the optimal trade-off between low electrical resistance and low threshold gain.

---

[a] Electronic mail: m.salih@leeds.ac.uk

## I. INTRODUCTION

Terahertz (THz) frequency quantum cascade lasers (QCLs)[1] have many potential applications across the physical and biological sciences (see, for example, Refs 2–6). However, the typical ~10-µm thickness of the THz QCL active region (AR) presents a number of challenges to its exploitation. First, the growth of such thick ARs by molecular beam epitaxy is technologically demanding, time consuming, and costly.[7] Second, high operating voltages are needed to achieve the required threshold electric fields, which can lead to undesirable device heating. This effect is particularly significant in high-performance QCLs based on resonant-phonon (RP) depopulation schemes,[8] which intrinsically require higher electrical input powers than chirped superlattice[1] or bound-to-continuum (BTC)[9] QCL designs. Higher operating biases are needed because of the relatively short AR module length and the need to incorporate a 36-meV LO-phonon emission within each period. Furthermore, there are typically large parasitic current channels in RP structures, which increases the threshold current density significantly.[10,11] As a result, continuous-wave operation is difficult to achieve. Nevertheless, the operating bias may be reduced through decreasing the device thickness. Although it is possible to obtain lasing from 1.75-µm ARs using double-metal (DM) plasmonic waveguides,[12] semi-insulating surface-plasmon (SP) waveguides are significantly easier to fabricate than DM structures, and intrinsically give much lower beam divergence without the need for additional complex processing steps. To date, thin SP waveguides have only been used with BTC designs, where an AR thickness of 5.86 µm was demonstrated.[7] In this work, we report the operation of high-performance RP QCLs with SP waveguide thicknesses ranging from 10 µm down to 5 µm. We compare the performance in terms of threshold current density, maximum operating temperature, output power and AR temperature as the AR thickness is reduced.

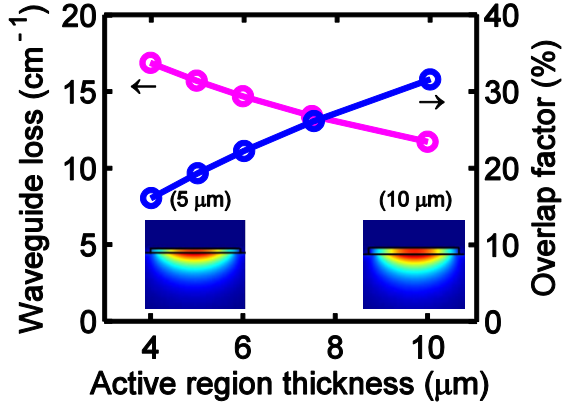

FIG. 1. (Color online) Calculated waveguide losses and overlap factors for SP waveguides with different AR thicknesses. The lines are intended only to guide the eye. Insets: two-dimensional waveguide modes calculated for 10 and 5-μm-thick ARs at 3.0 THz. The different colors represent the optical field intensity; with red the most intense, and blue the least intense. It was assumed that each device was 1.6-mm long and 150-μm wide; the upper n$^+$-GaAs layer was 80-nm-thick with doping level $N_d = 5 \times 10^{18}$ cm$^{-3}$, and the lower n$^+$-GaAs layer was 700-nm-thick with $N_d = 5 \times 10^{18}$ cm$^{-3}$. The gold over-layer was 150-nm thick, the GaAs substrate was 250-μm thick and the backside gold layer was 150 nm-thick.

## II.   THZ QCL FABRICATION AND CHARACTERIZATION

A series of four QCL wafers was grown by molecular beam epitaxy, based on a 3.1-THz GaAs/AlGaAs three-quantum-well resonant-phonon depopulation design[13] with AR thicknesses of 10, 7.5, 6 and 5 μm. The AR layer sequence in each case (starting from the injection barrier), with Al$_{0.15}$Ga$_{0.85}$As barriers indicated in bold, was **48**/96/**20**/74/**42**/53/<u>55</u>/53 Å. The uniformly doped layer ($N_d = 5 \times 10^{16}$ cm$^{-3}$) is underlined. The ARs were sandwiched between upper 80-nm-thick ($N_d = 5 \times 10^{18}$ cm$^{-3}$) and lower 700-nm-thick ($N_d = 5 \times 10^{18}$ cm$^{-3}$) GaAs contact layers. All devices were processed with SP waveguides[1] with ridge widths of 150 μm or 200 μm, and lengths in the range 1−3 mm. The thicknesses of the Au/Ge/Ni bottom and top contacts were 200 and 100 nm, respectively, and the thickness of the Ti/Au overlayer was 20/150 nm. Substrates were thinned to a thickness of 250 μm. For characterization, devices were mounted on the cold-finger of a continuous-flow cryostat equipped with polyethylene windows. Radiation was collected and coupled into a helium-cooled germanium bolometer using two off-axis parabolic reflectors, in a dry nitrogen-purged atmosphere. Emission spectra were acquired with a resolution of 7.5 GHz using a Bruker IFS66 FTIR spectrometer.

The lengths of the 10- and 7.5-µm-thick devices were nearly identical (1.53 and 1.55 mm, respectively), whereas 6-µm-thick devices of this length were found to operate with low output powers and low maximum operating temperatures (only ~15 K). In order to elucidate this observation, simulations of the SP waveguide were performed using a two-dimensional (2D) finite element model, with complex permittivities of the waveguide layers obtained from the bulk Drude model. Fig. 1 shows the calculated waveguide losses $\alpha_w$ and overlap factors $\Gamma$ of the optical mode with the AR for different AR thicknesses, and the insets show the calculated 2D waveguide modes for the cases of 10-µm and 5-µm-thick ARs. Decreasing the AR thickness leads to a decrease in $\Gamma$, as expected. Furthermore, despite a larger fraction of the mode overlapping with the undoped substrate, the waveguide losses are seen to increase in thinner devices owing to a larger overlap between the optical mode and the highly-doped bottom contact region. Nevertheless, this increased loss can be partly offset through reducing the effective facet losses by increasing the device length. Slightly longer devices, of length 1.67 and 1.83 mm, were therefore tested for the 6-µm-thick AR.

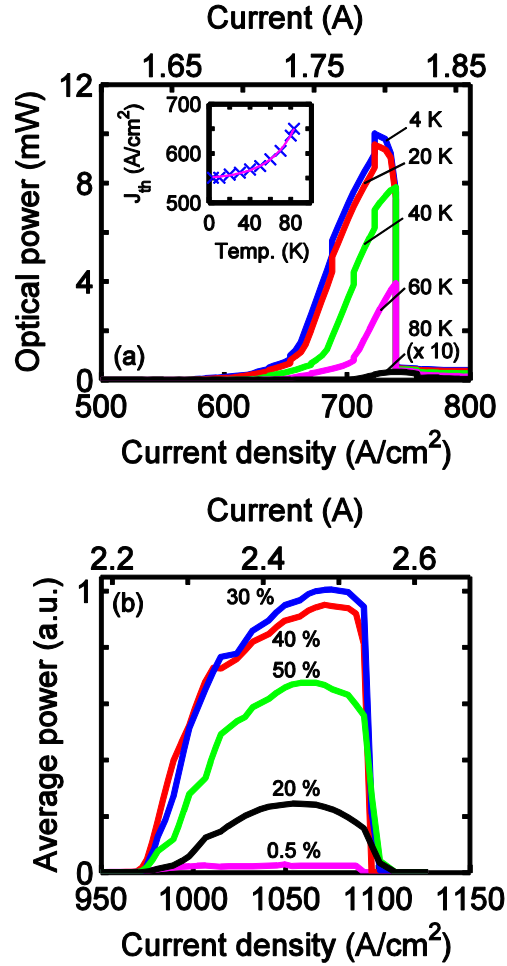

FIG. 2. (Color online) (a) Peak power–current density characteristics as a function of heat sink temperature for a 7.5-μm-thick AR driven at 2% duty-cycle at a repetition rate of 10 kHz, with the pulse train being electrically gated by a 167-Hz square-wave. Inset: variation of threshold current density with active-region temperature. The curve shows an exponential fit to the measured data. (b) Average power–current density characteristics as a function of duty-cycle for the 7.5-μm-thick device. The device was 1.55 mm long and 150 μm wide.

## III. RESULTS AND DISCUSSION

### A. Thermal analysis

The method described in Ref. 14 was used to characterize the thermal performance of each device. First, the relationship between the threshold current density $J_{th}$ and active region temperature $T_{AR}$ was determined, and fitted to the phenomenological exponential relation $J_{th} = J_0 + J_1\exp(T_{AR}/T_0)$, where $J_0$, $J_1$ and $T_0$ are fitting parameters. The QCL was driven at low duty-cycle to ensure that $T_{AR}$ was approximately equal to that of the cryostat cold finger. This was achieved using 1- or 2-µs pulses at a repetition rate of 10 kHz, with the pulse train being electrically gated by a 167-Hz square-wave, resulting in an effective duty-cycle ≤ 1%. Fig. 2(a) shows the peak output power measured as a function of current density at various heat sink temperatures for the 7.5-µm-thick device, when driven by 2-µs-long pulses. The inset shows the measured relationship between $J_{th}$ and $T_{AR}$. Equivalent relationships were obtained for the 6-, and 10-µm-thick devices.

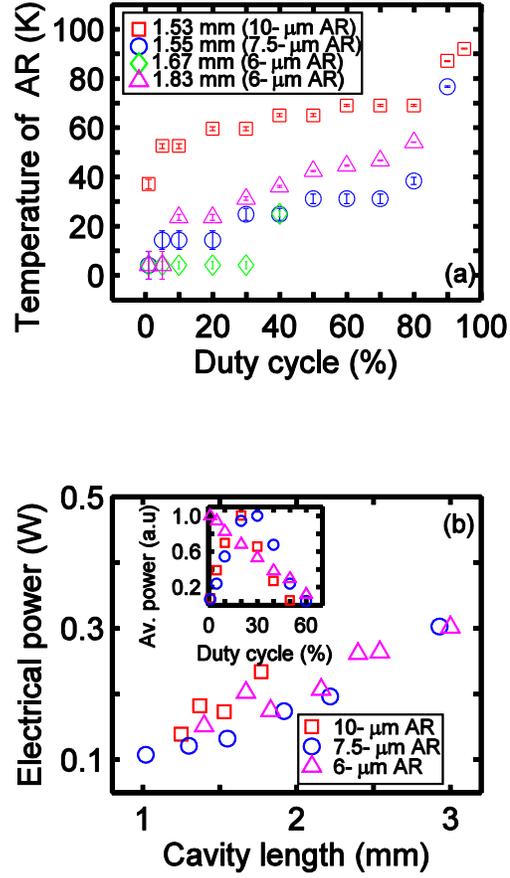

FIG. 3. (Color online) (a) Temperature of AR at threshold as a function of duty-cycle and, (b) variation of time-averaged electrical power at threshold as a function of cavity length, for 10, 7.5-, and 6-μm AR devices. Inset: normalized time-averaged power as a function of duty-cycle, for 10-μm-thick AR (device dimensions 1.53 mm × 150 μm), 7.5-μm-thick AR (1.53 mm × 150 μm), and 6-μm-thick AR (1.83 mm × 150 μm) devices.

The power–current-density relation was then measured for each device when driven by a continuous train of pulses with different duty-cycles, at a fixed repetition rate of 1 kHz. Fig. 2(b) shows the results for the 7.5-μm-thick AR with duty-cycles ranging from 0.5% to 50%. The temperature of the AR at threshold was then estimated for each duty-cycle through use of the phenomenological relationships between $J_{th}$ and $T_{AR}$, which were obtained from the low duty-cycle measurements shown in Fig. 2(a), as described above. Fig. 3(a) summarizes the relationship between duty-cycle and AR temperature at threshold for devices with AR thicknesses of 10, 7.5 and 6 µm. Overall, the results confirm that reducing the AR thickness leads to less electrical heating, owing to the reduced bias voltages required for operation, and potentially improved heat extraction from thinner devices. It is clear from Fig. 3(a) that despite the slightly longer lengths of the 6-µm-thick devices, lower AR temperatures are maintained when compared to the standard 10-µm active regions. Nevertheless, despite the reduced AR temperature in thinner devices, the largest duty-cycle at which devices lase is found to be limited by the progressively lower maximum operating temperature ($T_{max}$) as the AR thickness is reduced. Specifically, $T_{max}$ was measured to be 104, 83 and 30 K for the 10-µm, 7.5-µm and 6-µm (1.67-mm-long) devices, respectively. An increase in threshold current density from ~600 to ~820 Acm$^{-2}$ was also observed as the AR thickness decreased from 10 to 6 µm. The increased $J_{th}$ and reduced $T_{max}$ are attributed to the reduced AR overlap factor and increased waveguide losses in thinner devices. Indeed, the 2D finite element simulations indicate that the threshold gain increases from ~60 cm$^{-1}$ for the 10-µm-thick device to ~98 cm$^{-1}$ for the 6-µm-thick device, based on the complex propagation constants and confinement factors calculated for the waveguide modes (see Fig. 1) and accounting for facet losses. Significantly, however, the peak electrical power densities at threshold were measured to be 7.9, 5.4, and 8.1 kWcm$^{-2}$ for the 10-, 7.5- and 6-µm (1.67-mm-long) devices, respectively. A similar trend is seen in Fig. 3(b), which shows the peak electrical powers at threshold for 10-, 7.5- and 6-µm-thick devices of varying length. The marked reduction in power consumption for the 7.5-µm-thick devices, compared to the standard 10-µm devices, is principally a result of reduced operating bias and is accompanied by only a modest deterioration in $T_{max}$. For the 6-µm devices, however, the enhanced threshold current densities lead to larger electrical power densities at threshold than for the 7.5-µm devices, in spite of their comparable operating biases.

### B. Output power

The maximum time-averaged output power measured for three of the devices, when driven by a continuous train of pulses of different duty-cycles at a repetition rate of 1 kHz and with a heat-sink temperature of 4.2 K, is shown in the inset to Fig. 3(b). For clarity, the power is normalized for each given device; the average powers measured at a duty-cycle of 2% (20-µs pulses) were 60 µW, 8 µW and 60 nW for the 10, 7.5 and 6 µm (1.83-mm-long) AR devices, respectively. It can be seen from the inset to Fig. 3(b) that the time-averaged output power for the 7.5- and 10-µm ARs increases with respect to duty-cycle until a maximum time-averaged power is reached at optimal duty-cycles of 30% and 20%, respectively. The larger optimal duty-cycle for the 7.5-µm AR is attributed to the reduced electrical heating, which enables high peak powers to be maintained at higher duty-cycles. For the 6-µm AR, however, the average power decreases as the duty-cycle increases above 1%. In spite of the reduced AR temperatures in this case, the peak power drops rapidly with increasing duty-cycle as the AR temperature approaches $T_{max}$. The average output powers from the 10, 7.5 and 6-µm devices were found to increase to 500 µW, 210 µW and 30 µW respectively when driven by 2% duty-cycle (2-µs) pulses at the greater repetition rate of 10 kHz, as the shorter pulse lengths resulted in lower heating.

### C. Beam profiles

SP waveguides exhibit lower beam divergence than double-metal waveguides[15] and are generally favored for free-space THz imaging applications[16,17] and heterodyne mixing schemes[18,19] for applications including high-resolution gas spectroscopy.[20] Nevertheless, several approaches have been demonstrated to reduce the beam divergence of double-metal devices, including second-order[21] and third-order[22] distributed feedback gratings, coupling to horn antennas[23] and spoof surface plasmon structures,[24] although these approaches are technologically challenging.

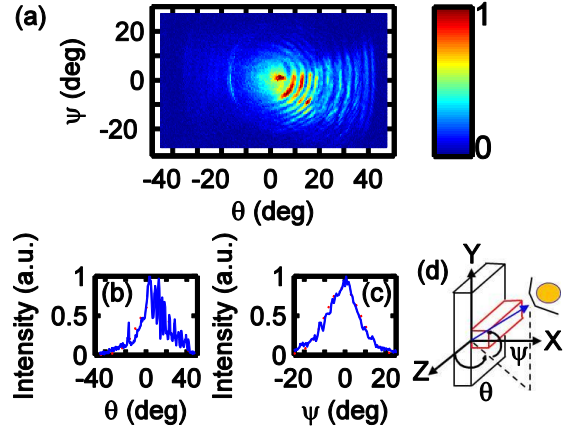

FIG. 4. (Color online) (a) Beam profile measured for a 6μm AR device of length 1.83 mm; (b) Beam profile cross section measured along the $\psi = 0°$ axis and (c) $\theta = 0°$ axis The dashed lines are Gaussian fits to the cross-sections. (d) Schematic of the laser, showing the coordinate system used in the measurements.

The far-field radiation patterns of devices were measured by raster-scanning a Golay cell with an aperture of 1-mm diameter, placed at a distance of 48 mm from the QCL, in order to examine the beam divergence in SP QCLs with reduced AR thickness. Fig. 4(a) shows the beam profile for the 6–μm device of length 1.83 mm. As has been described elsewhere[19] the radiation patterns can be understood in terms of two contributions: a broad envelope corresponding to a diffraction-limited beam generated at the QCL facet; and a ring-like interference pattern that can be understood using a wire laser model[25] that treats the QCL as a longitudinally-distributed source. From the beam profile cross sections measured along the $\psi = 0°$ and $\theta = 0°$ axes, which are shown in Fig. 4(b) and Fig. 4(c), the angular divergences of the beam (defined as the full-width at half-maximum of the angular distributions) are found to be 36° and 20°, respectively.

Fig. 4(d) shows the schematic of the laser, showing the co-ordinate system used in all measurements. For the 10–μm device, the values 37° and 25° are obtained. The similarity of the divergence for the 10-μm and 6-μm ARs can be explained by the fact that the ridge widths are equal and the waveguide mode extends similarly into the substrate in both cases (see Fig. 1), giving rise to similar diffraction-limited far-field distributions.

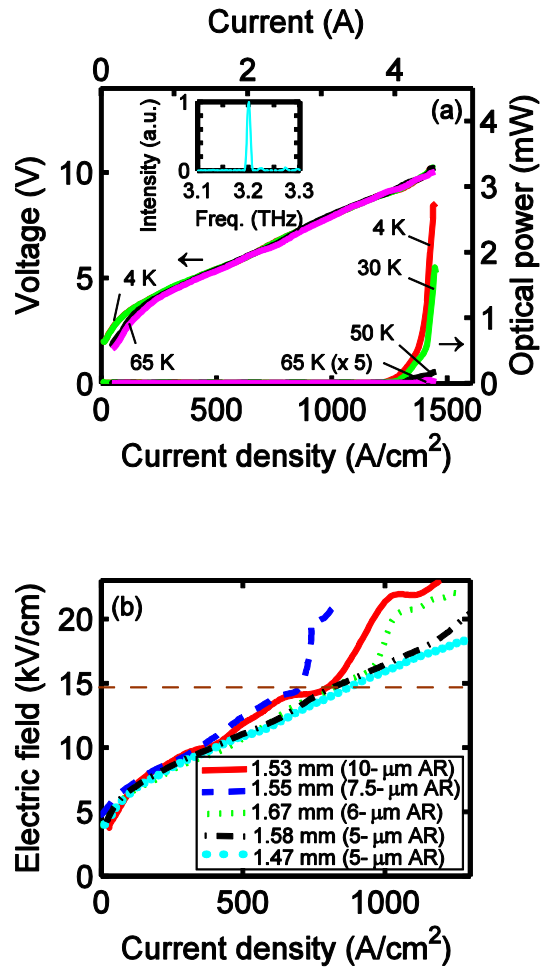

FIG. 5. (Color online) (a) Power–voltage–current density characteristics for a 5-μm AR device (1.58 mm × 200 μm). Inset: Emission spectrum of the same device. (b) Electric field–current density characteristics for 10-μm-thick (device dimensions 1.53 mm × 150 μm), 7.5-μm-thick (1.55 mm × 150 μm), 6-μm-thick (1.67 mm × 150 μm), and 5-μm-thick (1.58 mm × 200 μm and 1.47 mm × 200 μm) devices. The horizontal dotted line shows the field at which subband misalignment occurs in the 10-μm AR.

## IV. 5-μm-THICK ACTIVE REGIONS

Lasing was also achieved from 5-μm-thick devices (the thinnest SP THz QCLs demonstrated to date), although a wider ridge (200 μm) was required in this case. For duty-cycles greater than 1%, the threshold current density was beyond the limit of our available current source and therefore data for 5-μm devices is not included in the comparison shown in Fig. 3(a) and 3(b). Nevertheless, the power–voltage–current density characteristics of a 1.58 mm × 200 μm device with a 5-μm AR are shown in Fig. 5(a). This device lased up to a maximum heat-sink temperature of 66 K with up to 2.7 mW peak power at a heat-sink temperature of 4.2 K. The inset to Fig. 5(a) shows the emission spectrum of this device. Fig. 5(b) shows the electric field–current density characteristics for 10-, 7.5-, 6- and 5-μm AR devices. The dashed line indicates the start point of subband misalignment for the 10-μm AR. It can be seen that the characteristics are nearly identical until subband misalignment occurs, and that the field at which each device reaches subband misalignment is approximately the same for all ARs. As such, the dynamic range (i.e., the voltage range over which lasing is achieved) was ~6.6 V for the 10-μm AR, compared to ~3 V for the 6-μm AR.

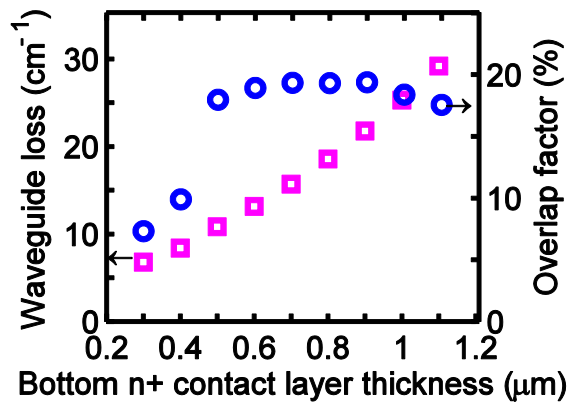

FIG. 6. (Color online) Calculated waveguide losses and overlap factors for 5-μm-thick SP waveguides with varying bottom n+ contact layer thickness. All other simulation parameters are identical to those in Fig. 1.

The performance of THz QCLs with thin ARs could potentially be improved though optimization of the SP waveguide structure. Specifically, it was found in section II that a significant fraction of the waveguide loss is caused by the overlap between the optical mode and the highly-doped bottom-contact layer. Fig. 6 shows finite-element simulation results for a 5-μm-thick SP waveguide with a range of bottom $n^+$ layer thicknesses. All other model parameters are identical to those used in Section II. The results show that structures with a contact-layer thicknesses of 500 nm or higher support a well-confined optical mode ($\Gamma \approx 18\%$). However, the simulated waveguide loss increases monotonically with the layer thickness. A 500-nm-thick contact layer therefore gives the best simulated performance with a 5-μm-thick AR: the waveguide loss is reduced from ~16 to ~11 cm$^{-1}$ as the layer thickness is reduced from 700 to 500 nm. Following this approach it may be possible to realize THz QCLs with active region thicknesses lower than 5 μm.

## V. CONCLUSION

We have demonstrated the operation of THz QCLs based on a three-well, resonant-phonon depopulation active region with a SP waveguide, for four different AR thicknesses (10, 7.5, 6 and 5 μm), the latter being the thinnest SP waveguide resonant-phonon QCL to date. Thinner ARs are technologically less demanding for MBE growth and result in reduced electrical heating of devices, but also lead to increased threshold currents and reduced maximum operating temperatures owing to increased waveguide losses and reduced overlap factors. Nevertheless, 7.5-μm-thick devices were found to give the lowest electrical power densities at threshold, indicating that they represent the optimal trade-off between low electrical resistance and low threshold gain in the AR. Furthermore, reducing the AR thickness was found to have minimal effect on the beam profile, thus confirming the suitability of such devices to a range of THz applications.

## VI. ACKNOWLEDGMENTS

We are grateful for the support of the EPSRC, the European Research Council programmes *NOTES* and *TOSCA*, the Royal Society and the Wolfson Foundation.